\documentclass[conference,10pt,letterpaper]{IEEEtran}%
\IEEEoverridecommandlockouts
\usepackage{amsmath}
\usepackage{graphicx}
\usepackage{multirow}
\usepackage[none]{hyphenat}
\usepackage{float}
\usepackage{subfig}
\usepackage{dblfloatfix}
\usepackage{cite}
\usepackage{t1enc}
\usepackage{times}
\makeatletter

\def\@IMSauthorblockNAMEstyle{\normalfont\IMSauthorsize}
\def\@IMSauthorblockAFFILstyle{\normalfont\IMSaffilsize}
\def\@IMSauthorblockEMAILstyle{\normalfont\IMSaffilsize}
\def\IMSauthorblockNAME#1{%
\relax\@IMSauthorblockNAMEstyle%
#1%
}%
\def\IMSauthorblockAFFIL#1{%
\relax\@IMSauthorblockAFFILstyle%
\vskip\@IEEEauthorblockAtopspace
#1%
}%
\def\IMSauthorblockEMAIL#1{%
\relax\@IMSauthorblockEMAILstyle%
\vskip\@IEEEauthorblockAtopspace
#1%
}%
\newcommand{\IMSauthor}[1]{%
\ifIsBlindReviewVersion%
\author{\phantom{\parbox{\textwidth}{\center\relax#1}}}%
\else%
\author{\parbox{\textwidth}{\center\relax#1}}%
\fi%
}%
\newif\ifIsBlindReviewVersion

\def\IMSthispaperforfinalpublication{\IsBlindReviewVersionfalse}
\def\@maketitle{\newpage
\bgroup\par\addvspace{0.5\baselineskip}\centering%
\ifCLASSOPTIONtechnote
   {\bfseries\large\@IEEEcompsoconly{\sffamily}\@title\par}\vskip 1.3em{\lineskip .5em\@IEEEcompsoconly{\sffamily}\@author
   \@IEEEspecialpapernotice\par{\@IEEEcompsoconly{\vskip 1.5em\relax
   \@IEEEtitleabstractindextextbox{\@IEEEtitleabstractindextext}\par
   \hfill\@IEEEcompsocdiamondline\hfill\hbox{}\par}}}\relax
\else
   \vskip0.2em{\IMStitlesize\ifCLASSOPTIONtransmag\bfseries\LARGE\fi\@IEEEcompsoconly{\sffamily}\@IEEEcompsocconfonly{\normalfont\normalsize\vskip 2\@IEEEnormalsizeunitybaselineskip
   \bfseries\Large}\@title\par}\vskip1.0em\par
   \ifCLASSOPTIONconference%
      {\@IEEEspecialpapernotice\mbox{}\vskip\@IEEEauthorblockconfadjspace%
       \mbox{}\hfill\begin{@IEEEauthorhalign}\@author\end{@IEEEauthorhalign}\hfill\mbox{}\par}\relax
   \else
      \ifCLASSOPTIONpeerreviewca
         {\@IEEEcompsoconly{\sffamily}\@IEEEspecialpapernotice\mbox{}\vskip\@IEEEauthorblockconfadjspace%
          \mbox{}\hfill\begin{@IEEEauthorhalign}\@author\end{@IEEEauthorhalign}\hfill\mbox{}\par
          {\@IEEEcompsoconly{\vskip 1.5em\relax
           \@IEEEtitleabstractindextextbox{\@IEEEtitleabstractindextext}\par\hfill
           \@IEEEcompsocdiamondline\hfill\hbox{}\par}}}\relax
      \else
         \ifCLASSOPTIONtransmag
           {\@IEEEspecialpapernotice\mbox{}\vskip\@IEEEauthorblockconfadjspace%
            \mbox{}\hfill\begin{@IEEEauthorhalign}\@author\end{@IEEEauthorhalign}\hfill\mbox{}\par
           {\vspace{0.5\baselineskip}\relax\@IEEEtitleabstractindextextbox{\@IEEEtitleabstractindextext}\vspace{-1\baselineskip}\par}}\relax
         \else
           {\lineskip.5em\@IEEEcompsoconly{\sffamily}\sublargesize\@author\@IEEEspecialpapernotice\par
           {\@IEEEcompsoconly{\vskip 1.5em\relax
            \@IEEEtitleabstractindextextbox{\@IEEEtitleabstractindextext}\par\hfill
            \@IEEEcompsocdiamondline\hfill\hbox{}\par}}}\relax
         \fi
      \fi
   \fi
\fi\par\addvspace{0.0\baselineskip}\egroup}

\def\IMStitlesize{\@setfontsize{\IMStitlesize}{18}{21pt}}
\def\IMSauthorsize{\@setfontsize{\IMSauthorsize}{12}{13pt}}
\def\IMSaffilsize{\@setfontsize{\IMSaffilsize}{12}{13pt}}
\def\IMScaptionsize{\@setfontsize{\IMScaptionsize}{8}{9pt}}
\def\IMSbibsize{\@setfontsize{\IMSbibsize}{8}{9pt}}

\def\@IEEEauthorblockNstyle{\IMSauthorsize\@IEEEcompsocnotconfonly{\sffamily}\@IEEEcompsocconfonly{\large}}
\def\@IEEEauthorblockAstyle{\IMSaffilsize\@IEEEcompsocnotconfonly{\sffamily}\@IEEEcompsocconfonly{\itshape}\@IEEEcompsocconfonly{\large}}
\def\@IEEEauthordefaulttextstyle{\IMSauthorsize\@IEEEcompsocnotconfonly{\sffamily}\sublargesize}

\def\thebibliography#1{\section*{\refname}%
    \addcontentsline{toc}{section}{\refname}%
    \IMSbibsize\@IEEEcompsocconfonly{\small}\vskip 0.3\baselineskip plus 0.1\baselineskip minus 0.1\baselineskip
    \list{\@biblabel{\@arabic\c@enumiv}}%
    {\settowidth\labelwidth{\@biblabel{#1}}%
    \leftmargin\labelwidth
    \advance\leftmargin\labelsep\relax
    \itemsep \IEEEbibitemsep\relax
    \usecounter{enumiv}%
    \let\p@enumiv\@empty
    \renewcommand\theenumiv{\@arabic\c@enumiv}}%
    \let\@IEEElatexbibitem\bibitem%
    \def\bibitem{\@IEEEbibitemprefix\@IEEElatexbibitem}%
\def\newblock{\hskip .11em plus .33em minus .07em}%
\ifCLASSOPTIONtechnote\sloppy\clubpenalty4000\widowpenalty4000\interlinepenalty100%
\else\sloppy\clubpenalty4000\widowpenalty4000\interlinepenalty500\fi%
    \sfcode`\.=1000\relax}

\long\def\@makecaption#1#2{%
\ifx\@captype\@IEEEtablestring%
\par\@IEEEtabletopskipstrut
\else
\@IEEEfigurecaptionsepspace
\fi
\setbox\@tempboxa\hbox{\normalfont\IMScaptionsize {#1.}\nobreakspace\nobreakspace #2}%
\ifdim \wd\@tempboxa >\hsize%
\setbox\@tempboxa\hbox{\normalfont\IMScaptionsize {#1.}\nobreakspace\nobreakspace}%
\parbox[t]{\hsize}{\normalfont\IMScaptionsize\noindent\unhbox\@tempboxa#2}%
\else
\ifCLASSOPTIONconference \hbox to\hsize{\normalfont\IMScaptionsize\hfil\box\@tempboxa\hfil}%
\else \hbox to\hsize{\normalfont\IMScaptionsize\box\@tempboxa\hfil}%
\fi\fi
\ifx\@captype\@IEEEtablestring%
\@IEEEtablecaptionsepspace
\else
\fi}

\newlength\tablecaptiontotableskip
\newlength\figuretocaptionskip
\def\@IEEEfigurecaptionsepspace{\vskip\figuretocaptionskip\relax}%
\def\@IEEEtablecaptionsepspace{\vskip\tablecaptiontotableskip\relax}%

\def\abstract{\normalfont%
\@IEEEabskeysecsize\bfseries\textit{\abstractname}\,\bfseries\textit{---}\,%
\@IEEEgobbleleadPARNLSP}%

\def\IEEEkeywords{\normalfont%
\@IEEEabskeysecsize\bfseries\textit{\IEEEkeywordsname}\,\bfseries\textit{---}\,%
\@IEEEgobbleleadPARNLSP}%
\def\endIEEEkeywords{\relax\vspace{0.67ex}%
\par\if@twocolumn\else\endquotation\fi%
\normalsize\normalfont}%

\DeclareRobustCommand*{\IMSauthorrefmark}[1]{\raisebox{0pt}[0pt][0pt]{\textsuperscript{\footnotesize{#1}}}}%
\def\@IEEEauthorblockNtopspace{0ex}
\def\@IEEEauthorblockAtopspace{1mm}
\def\IEEEkeywordsname{Keywords}
\def\subsubsection{\@startsection{subsubsection}{3}{\z@}{1.5ex plus 1.5ex minus 0.5ex}%
{0.7ex plus .5ex minus 0ex}{\normalfont\normalsize\itshape}}%
\def\@seccntformat#1{\csname the#1dis\endcsname\relax}
\def\thesubsectiondis{{\hbox to\parindent{\Alph{subsection}.}}}
\def\thesubsubsectiondis{{\hbox to \parindent{\arabic{subsubsection})}}}
\def\theparagraphdis{{\hbox to \parindent{\alph{paragraph})}}}
\IEEEilabelindentA \parindent
\IEEEilabelindent \IEEEilabelindentA
\IEEEelabelindent \parindent
\IEEEdlabelindent \parindent
\IEEElabelindent \parindent

\newlength\@IMSparindent
\newcommand\IMSdisplayacksection[1]{%
\ifIsBlindReviewVersion%
\noindent\phantom{\parbox[t]{\columnwidth}{\normalbaselines\setlength{\parindent}{\@IMSparindent}{#1}\strut}}
\else%
\noindent\parbox[t]{\columnwidth}{\normalbaselines\setlength{\parindent}{\@IMSparindent}{#1}\strut}%
\fi%
}%

\makeatother

\usepackage{fancyhdr}
\fancypagestyle{firststyle}{\fancyhf{}
	\fancyhead[L]{\small D. Shakya, T. S. Rappaport, E. Shieh, M. E. Knox, H. Rahmani, D. Shahrjerdi, M. Ying, K. Fan, M. Lu, A. Rumiantsev, V. Mallette, G. Fisher, G. De Chirico, P. Ghate, and S. McMahon, ``Four-Port Probe Stations and SOLR Calibration Standard Design up to 125 GHz on 28 nm CMOS",\textit{ (Accepted) IEEE Asia Pacific Microwave Conference, }South Korea, Dec. 2025, pp. 1--3.}
}

\begin{document}
	\bstctlcite{IEEEtran:BSTcontrol}
	\raggedbottom
	%
	%
	%
	\title{Four-Port Probe Stations and SOLR Calibration Standard Design up to 125 GHz on 28 nm CMOS
		\thanks{This work is supported by NSF MRI grant 2216332 and NYU WIRELESS Industrial Affiliates Program}
	}
	
	%
	%
	%
	\IMSthispaperforfinalpublication
	\IMSauthor{%
		\IMSauthorblockNAME{
			Dipankar Shakya\IMSauthorrefmark{\#1},
			Theodore S. Rappaport\IMSauthorrefmark{\#2},
			Ethan Shieh\IMSauthorrefmark{\#3},
			Michael E. Knox\IMSauthorrefmark{\#4},
			Hamed Rahmani\IMSauthorrefmark{\#5},
			Davood Shahrjerdi\IMSauthorrefmark{\#6},
			Mingjun Ying\IMSauthorrefmark{\#7},
			Kimberley Fan\IMSauthorrefmark{*1},
			Matt Lu\IMSauthorrefmark{*2}, 
			Andrej Rumiantsev\IMSauthorrefmark{*3},\\ 
			Vince Mallette\IMSauthorrefmark{*4},
			Gavin Fisher\IMSauthorrefmark{\textasciicircum1},Giancarlo De Chirico\IMSauthorrefmark{\textasciicircum2}, Pratik Ghate\IMSauthorrefmark{\textasciicircum3}, Shean McMahon\IMSauthorrefmark{\textasciicircum4}
		}
		\\%
		\IMSauthorblockAFFIL{
			\IMSauthorrefmark{\#}NYU WIRELESS, New York University, USA\\
			\IMSauthorrefmark{*}MPI Corporation\\
			\IMSauthorrefmark{\textasciicircum}FormFactor Incorporated
		}
		\\%
		\IMSauthorblockEMAIL{
			\{\IMSauthorrefmark{\#1}dshakya, 
			\IMSauthorrefmark{\#2}tsr,
			\IMSauthorrefmark{\#3}es5185, 
			\IMSauthorrefmark{\#4}mikeknox,
			\IMSauthorrefmark{\#5}hr926,
			\IMSauthorrefmark{\#6}davood\}@nyu.edu
		}
	}
	%
	\maketitle
	
	\thispagestyle{firststyle}
	%
	%
	%
	\begin{abstract}
		This paper presents two innovative four-port probe stations developed by FormFactor Incorporated (FFI) and MPI Corporation (MPI), and a four-port calibration standard design up to 125 GHz for the probe stations. True four-port probing at mmWave and beyond does not yet exist, but is anticipated for future multi-band wireless devices using several antennas and RF chains. The four-port probe stations are housed in the THz measurement facility at NYU and allow simultaneous probing from East, West, North, and South orientations, which presents challenges for calibration. An on-chip Short-Open-Load-Reciprocal (SOLR) calibration (cal) standard is designed leveraging UMC's 28 nm CMOS process. S/O/L standard S-parameters are extracted using a virtual multiline Thru-Reflect-Line (mTRL) cal and used to validate SOLR cal performance via simulations up to 125 GHz. The novel probing solutions from MPI and FFI, along with the SOLR cal, open up considerable opportunities for precise RF characterization across wide frequency ranges.
	\end{abstract}
	%
	%
	
	\section{Introduction}
	In the rapidly evolving landscape of wireless communication, the millimeter-wave (mmW) (30-300 GHz), sub-THz (100-300 GHz) and THz (0.3-3 THz) spectrum presents a new frontier with the potential to enhance the capacity, speed, and diversity of wireless applications \cite{Rappaport2019ia}. 
	At these mmW and THz bands, where short wavelengths and modern semiconductor nodes support massive numbers of antennas and RF chains, including devices such as multi-band multi-output mixers and amplifiers, advanced equipment calibration (cal) and measurement techniques are essential. This paper presents the first commercial four-port probe stations for simultaneous measurements across four ports from DC to 500 GHz established at the THz Measurement facility in NYU WIRELESS\cite{Shakya2024IMM}. Moreover, a novel SOLR cal design to enable measurements on the four-port probe station up to 125 GHz is presented.
	
	
	Traditionally, two or three-port RF probe stations have sufficed for on-chip measurements. However, advances in wireless communications provide compelling reasons for the widespread adoption of four-port probe stations. Particularly, multiple-input multiple-output (MIMO) transceivers and multi-band carrier aggregation systems, can contain several RF chains that require simultaneous probing to assess phase shifts and inter-chain leakage/coupling \cite{Shakya2024IMM}. Four-port probing allows faster and more accurate simultaneous measurements between RF chains without repositioning probes. Moreover, with four-port probing, devices such as couplers and quadrature mixers are characterized in a single measurement with complete four-port S-parameters captured.
	
	Several Vector Network Analyzer (VNA) cal techniques have been developed to accurately set the measurement reference plane for traditional two-port probe station measurements. Short-Open-Load-Thru (SOLT) and Thru-Reflect-Line (TRL) cal using off-chip impedance standard substrates (ISS) are considered reliable at lower frequencies (typically below 20 GHz\cite{stenarson2007arftg}). mTRL is commonly adopted to extend the two-port cal up to hundreds of GHz leveraging the redundancy of multiple Line standards\cite{rumiantsev2008arftg,Krozer2014bctm}. However, the SOLR cal for orthogonal ports, though practiced for several years\cite{Basu1997ims}, have been limited to below 26 GHz\cite{Daniel2005USF}. 
	
	Paper organization is as follows: Section \ref{sxn:4portPS} introduces the two four-port probe stations developed independently by MPI and FFI. Section \ref{sxn:Cal} discusses an SOLR cal design up to 125 GHz in UMC 28 nm CMOS. Section \ref{sxn:Res} presents the SOLR cal performance before concluding. 
	
	
	\section{The Four-Port Probe Station}
	\label{sxn:4portPS}
	
	\subsection*{The THz Measurement Facility (THz Lab)}
	The \textit{THz Lab} is a multiuser lab at NYU WIRELESS established by the National Science Foundation (NSF) in 2022 that enables measurements of devices, circuits, materials, and radio channels from DC to 500 GHz \cite{Shakya2024IMM}. 
	
	
	The \textit{THz Lab} employs a Keysight N5247B PNA-X network analyzer with a N5292A test-set controller to enable four-port S-parameter measurements. 
	Banded waveguide frequency extenders from Virginia Diodes, Inc. (VDi) are used with the PNA-X to enable frequency-swept measurements from DC to 500 GHz. 
	FFI's Infinity waveguide probes with a ground-signal-ground (GSG) configuration are used in the \textit{THz Lab} for the VDi frequency extenders \cite{rumiantsev2013imm,Shakya2024IMM}. 
	
	\subsection{The FormFactor Four-Port Probe Station}
\begin{figure}[htbp]
		\centering
		\includegraphics[width= 60 mm, height= 48 mm]{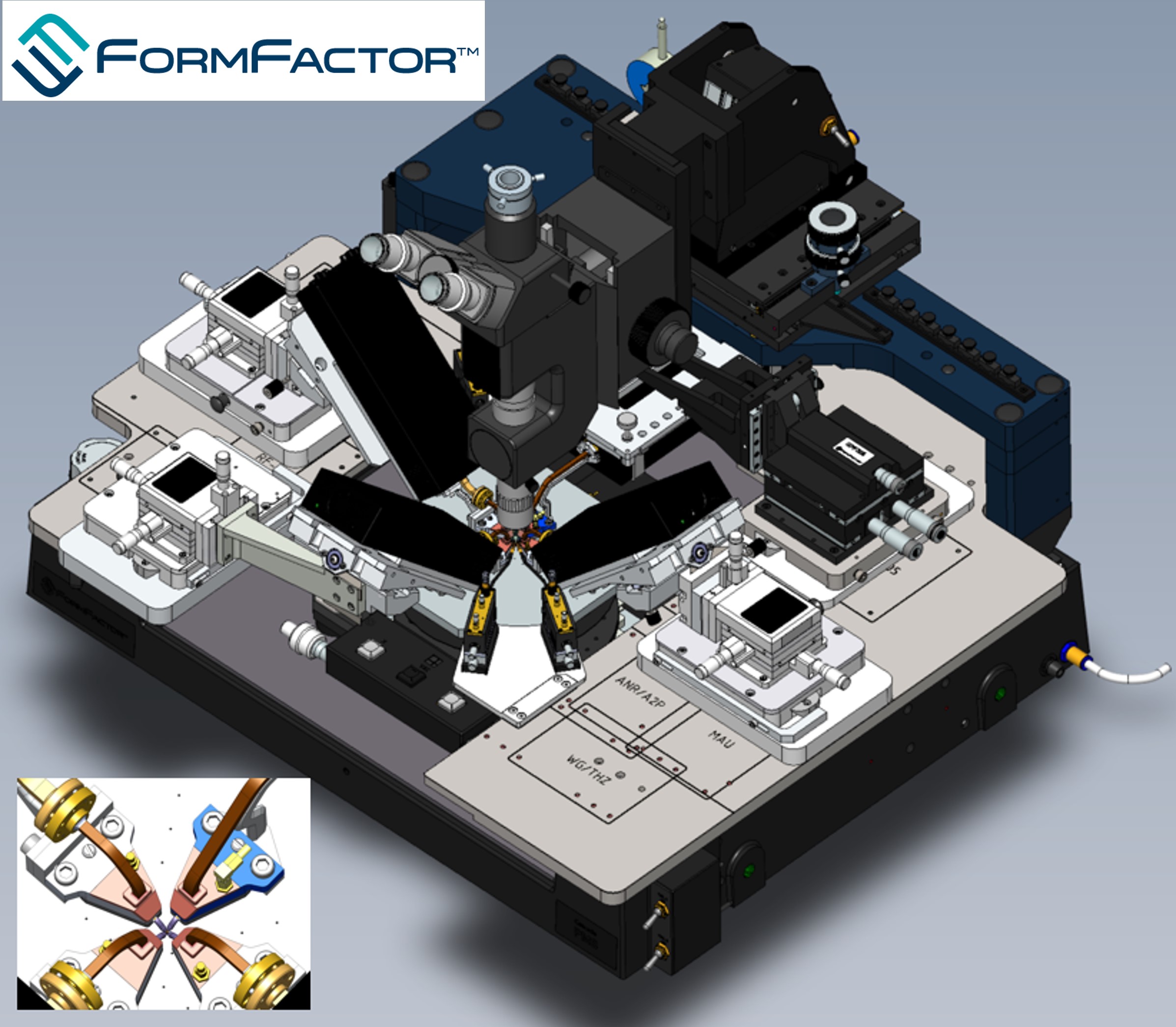}
		\vspace{-2 pt}
		\caption{Formfactor Inc. design of the four-port probing solution. Design includes three angled positioners and one flat North positioner.}
		\label{fig:FFI4port}
		\vspace{-10 pt}
	\end{figure}

	The FFI manual four-port probe station is engineered based on the EPS-200 mmW probing platform (Fig. \ref{fig:FFI4port}). This station incorporates a synthesis of FFI's distinctive 45-degree angled positioners on the East, West, and South orientations, with the traditional flat positioner in the North. To accommodate the North positioner and frequency extender, the bridge supporting the microscope is elevated. 
	The station utilizes a vacuum-based chuck incorporating movement control mechanisms that permit constrained linear motion in either the X or Y axes, or unrestricted movement across both axes.
	
	\subsection{The MPI Four-Port Probe Station}
\begin{figure}[htbp]
		\centering
		\vspace{-5 pt}
		\includegraphics[width=62 mm, height= 48 mm]{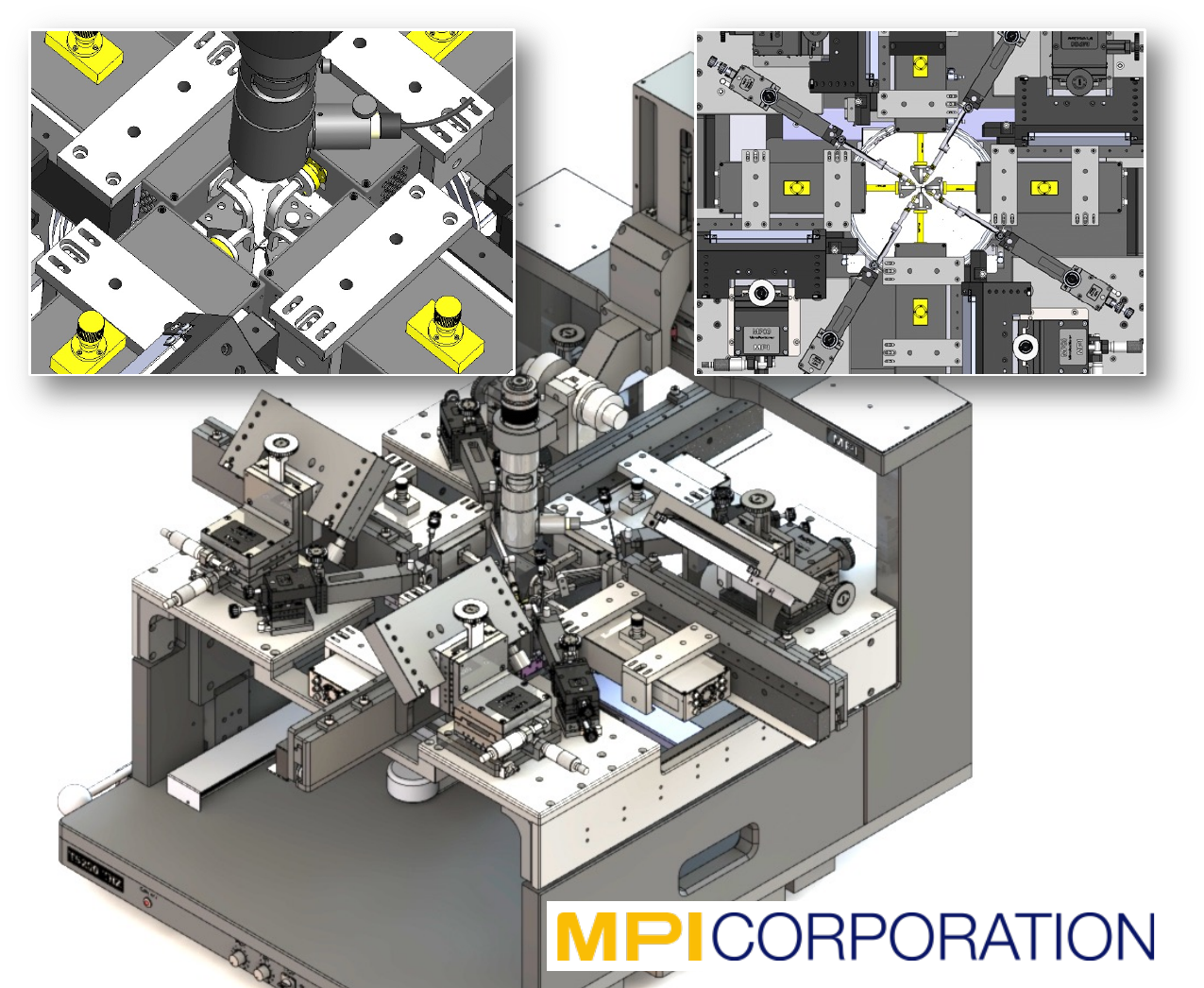}
		\vspace{-2 pt}
		\caption{MPI Corporation design for the four-port probing platform. Zoomed-in views show direct mounting of waveguide probes onto frequency extenders.}
		\label{fig:MPI4port}
		\vspace{-10 pt}
	\end{figure}

	The MPI design evolves from the TS200-THZ 200 mm mmW, THz, and load pull probe station (Fig. \ref{fig:MPI4port}). This design allows direct connection of any VDi frequency extender to the probe across all four orientations: East/West, or North/South, and eliminates S-bends or additional waveguides, significantly reducing losses and maximizing the signal power and system directivity. 
	The scope bridge is raised for ease of access to controls. The MPI air-bearing stage, with single-handed puck control, enables fast unrestricted XY navigation and quick wafer loading. 
	Additionally, a safety lock restricts all chuck movement at the platen's lowest probe contact position.
	
	\section{Calibration of the four-port probe stations}
	\label{sxn:Cal}
	
	A critical aspect for the four-port probe station implementation is the calibration of orthogonal ports such as the North-East or South-West. Difficulty arises due to the requirement of a non-ideal Thru standard to connect the orthogonal ports during cal. 
	The non-ideal Thru inhibits useful application of any error correction technique that needs well-defined Thru or Line standards, such as SOLT or TRL \cite{rumiantsev2022crc,Basu1997ims}.
	
	\subsubsection{Viability of SOLR calibration}
	The SOLR methodology for on-wafer cal appears as a viable approach for calibrating orthogonal port configurations \cite{Basu1997ims}. The SOLR requires three fully known one-port standards at each port, namely, Short-Open-Load (SOL), with one unknown reciprocal standard between any two VNA ports to determine error terms of the underlying eight-term error model\cite{rytting2000arftg}. The reciprocity, S12 = S21, and low signal loss are the only requirements for the reciprocal standard\cite{shoaib2017sp}. 
	The calibration quality is subject to the quality of the one-port standard definitions available to the VNA \cite{shoaib2017sp, rumiantsev2008vna}. These standards, however, could be characterized with a polynomial fit established via alternative cal like the mTRL derived from straight Thru standards \cite{Arz2023arftg}. 
	
	\subsubsection{Calibration Standard Design on UMC 28 nm CMOS}
	Advanced semiconductor processes such as CMOS and RF-SOI, at sub-micron technology nodes (e.g., 130 nm or 28 nm) are preferred for fabricating mmW and sub-THz transceivers considering cost, \textit{fT/fmax}, noise performance, etc. \cite{Shakya2024IMM,rumiantsev2008arftg}.
	While off-chip ISS are available to implement various cal techniques, 
	on-chip cal standards fabricated on the same wafer or die as the device achieve the highest accuracy and repeatability, especially as frequencies extend into mmW and sub-THz range. On-chip standards allow the measurement reference plane to be moved closest to the device, while also accounting for influences of the process dielectric stackup\cite{rumiantsev2022crc}. 
	
	Calibration standards for an SOLR cal are taped-out on the UMC 28 nm CMOS process\cite{UMC2021rfcmos}. A nine metal layer stackup with Aluminum (Al) landing pads is used for the 4 mm $\times$ 4 mm die size. Calibration structures are fabricated on the top metal layer (M9) and a composite ground with M2 and M1 at the bottom is used. 
	Simulations are carried out replicating the stackup in Ansys HFSS as per UMC documentation. 
	
\begin{figure}[htbp]
		\centering
		\vspace{-5 pt}
		\includegraphics[width=0.41\textwidth]{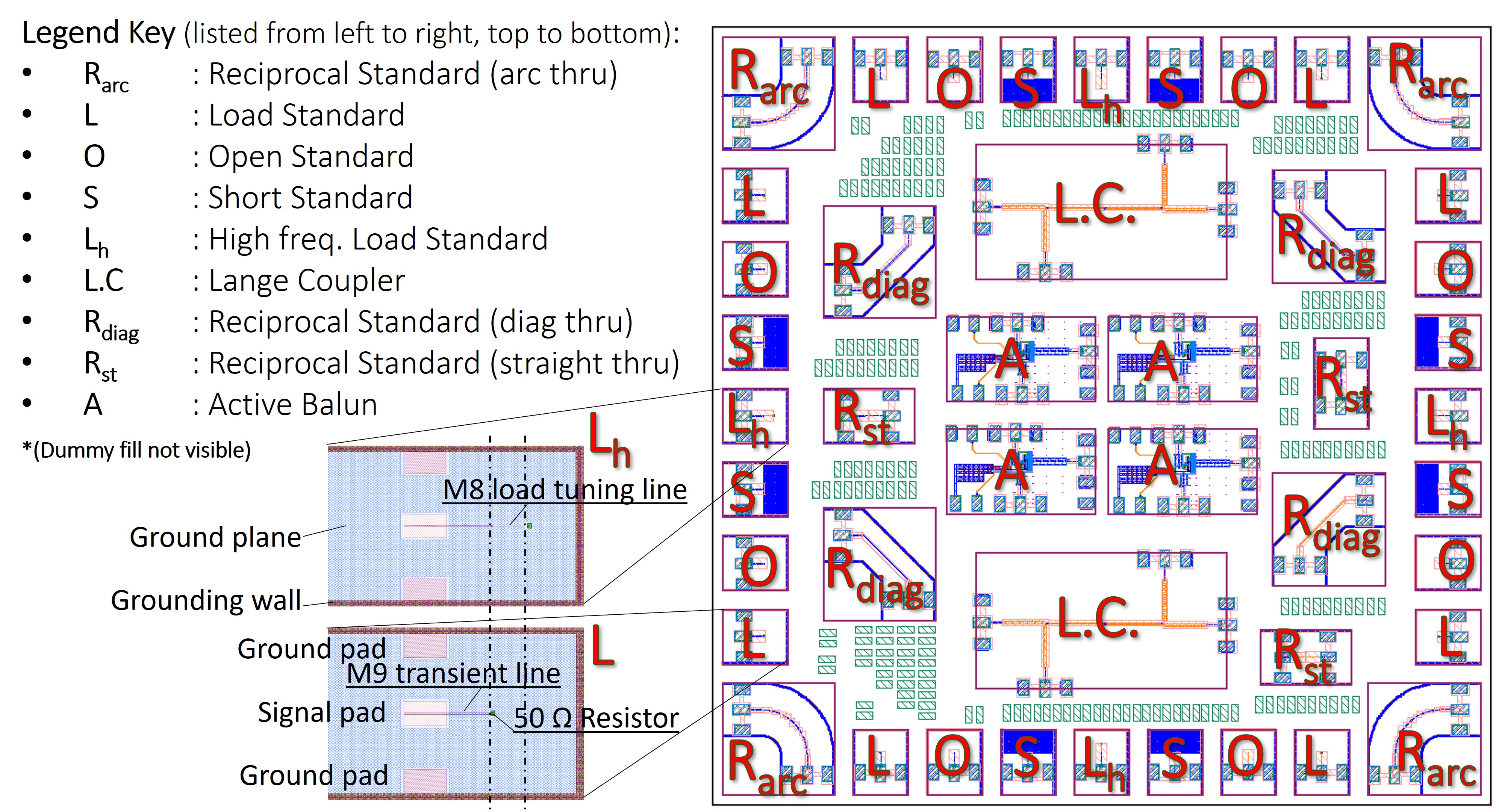}
		\vspace{-2 pt}
		\caption{Simplified layout of the taped-out 4 mm $\times$ 4 mm SOLR cal chip. SOL standards line the die edges. Arced Thru (R) standard at corners. Diagonal and straight Thrus at chip center. Four SOLR cal set replicas included}
		\label{fig:CalChip}
		\vspace{-15 pt}
	\end{figure}

	Pad size and geometry are optimized (35 $\mu$m $\times$ 55 $\mu$m) to minimize parasitic capacitance and resistance for GSG probing, while adhering to FFI and UMC design rules\cite{rumiantsev2022crc,wang2019cstic}. 
	The taped-out die is shown in Fig. \ref{fig:CalChip} with one-port SOL standards along each die edge and Reciprocal orthogonal Thrus at the corners. Two types of orthogonal Thrus are implemented: an arc to avoid sharp bends, and diagonal with 45$^\circ$ bends. Total four replica sets of the SOLR cal, including both arc and diagonal Thrus, are fabricated on the die.  
	
	A microstrip line with an optimized length (55 $\mu$m) and width (2 $\mu$m) on M9 is implemented between the pad and each standard to maximize the calibration frequency range. A 50 $\Omega$ poly-silicon metal gate resistor serves as the Load standard, with an additional 45 $\mu$m tuning line on M8 improving its performance. Consistent M9 microstrip lines on all standards move the measurement plane at the line's end after calibration. Reciprocal standards are implemented using straight, arc, and diagonal Thru lines for connecting orthogonal ports.
	
	
	
	\section{Performance Results of the SOLR cal}
	\label{sxn:Res}
\begin{figure}[htbp]
		\centering%
		\vspace{-10pt}
		\subfloat[]{%
			\centering
			\includegraphics[width=40mm]{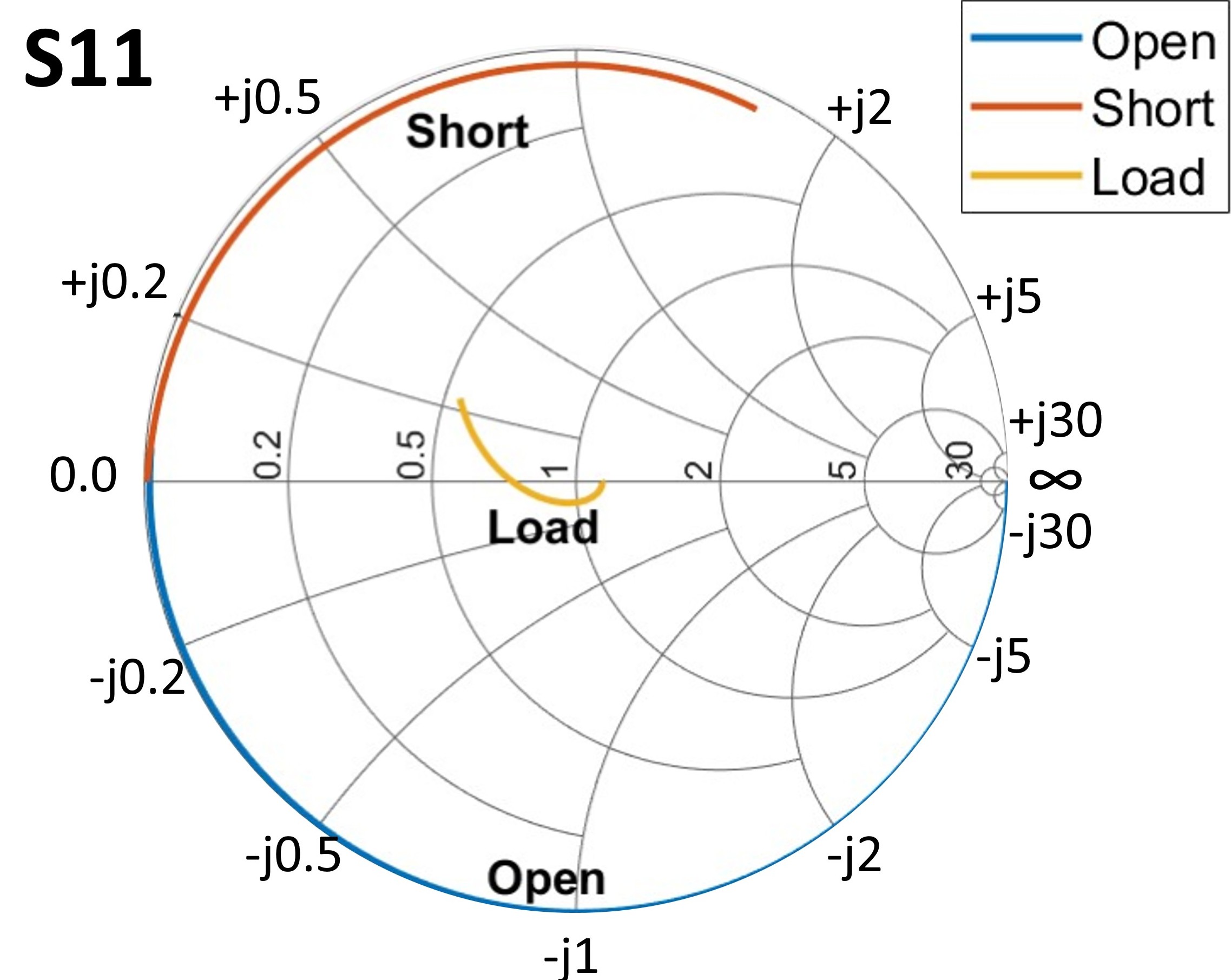}
			\label{fig:lores-photo}
		}%
		\hspace{-5pt}
		\subfloat[]{%
			\centering
			\includegraphics[width=45mm]{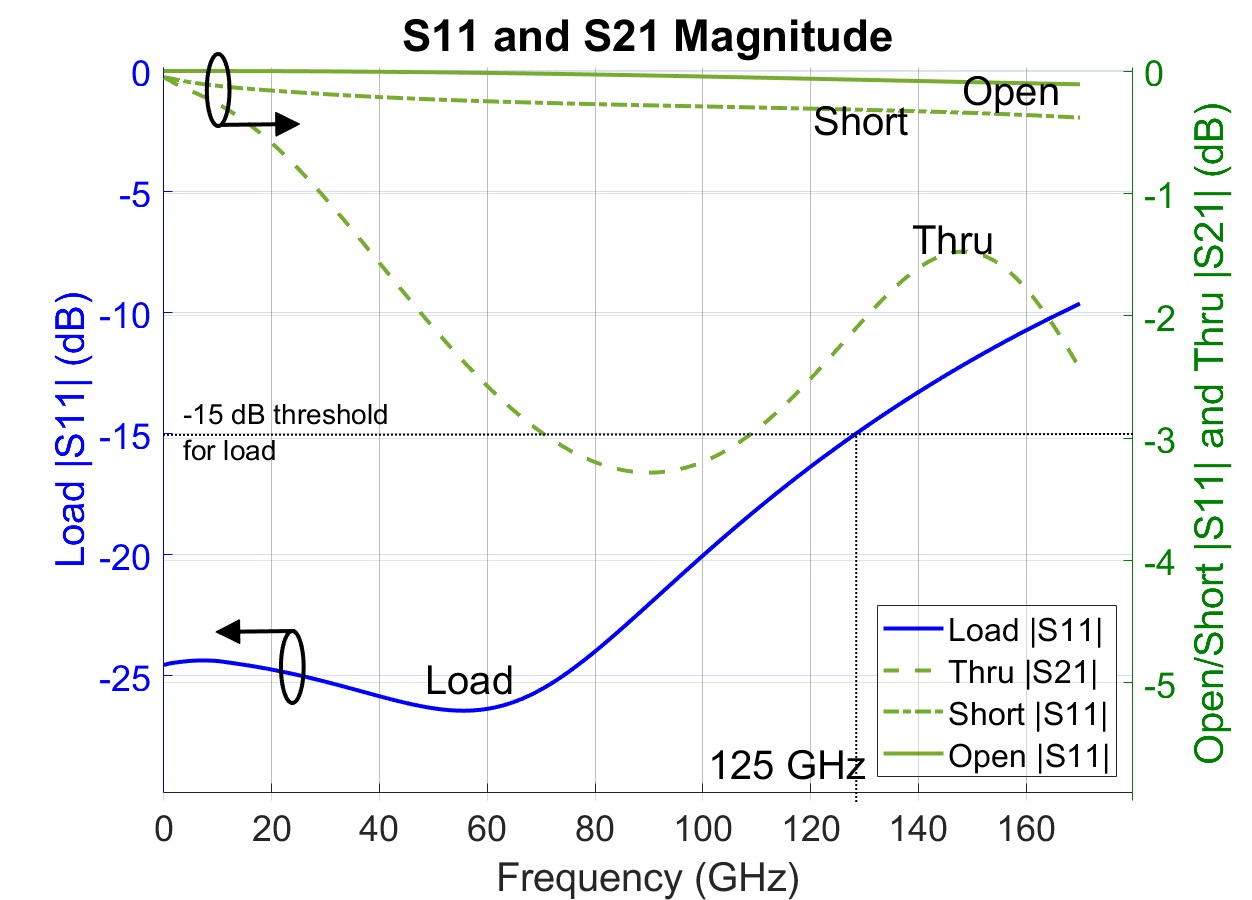}
			\label{fig:hires-photo}
		}%
		\vspace{2pt}
		\caption{Simulated plots of s-parameters for the SOLR cal: (a) $S_{11}$ impedance of Open, Short, and Load; (b) $|S_{11}|$ of Load (left Y-axis); $|S_{11}|$ of Open and Short, $|S_{21}|$ of Orthogonal Thru (right Y-axis).}
		\label{fig:StdRes}
		\vspace{-5pt}
	\end{figure}

	The SOLR cal standards for four-port probe station calibration simulated in HFSS demonstrate robust performance up to 125 GHz. The Open standard consistently maintains an $|S_{11}|$ response above -0.18 dB up to 170 GHz, with capacitive arcing behavior on the Smith chart as frequency increases. Similarly, Short standard $|S_{11}|$ response remains above -0.7 dB up to 170 GHz, showing the anticipated inductive arc on the Smith chart (Fig. \ref{fig:StdRes}). The phase response of the two standards shows $\sim$180$^\circ$ phase difference across D-band.
	
	The Load standard exhibits an $|S_{11}|$ response below -15 dB up to 125 GHz, which is employed as the threshold for establishing cal performance \cite{Daniel2005USF}. 
	All three Thru lines consistently meet the reciprocity requirement ($|S_{21}|=|S_{12}|$). 
	
\begin{figure}[htbp]
		\centering%
		\vspace{-10pt}
		\includegraphics[width=0.46\textwidth]{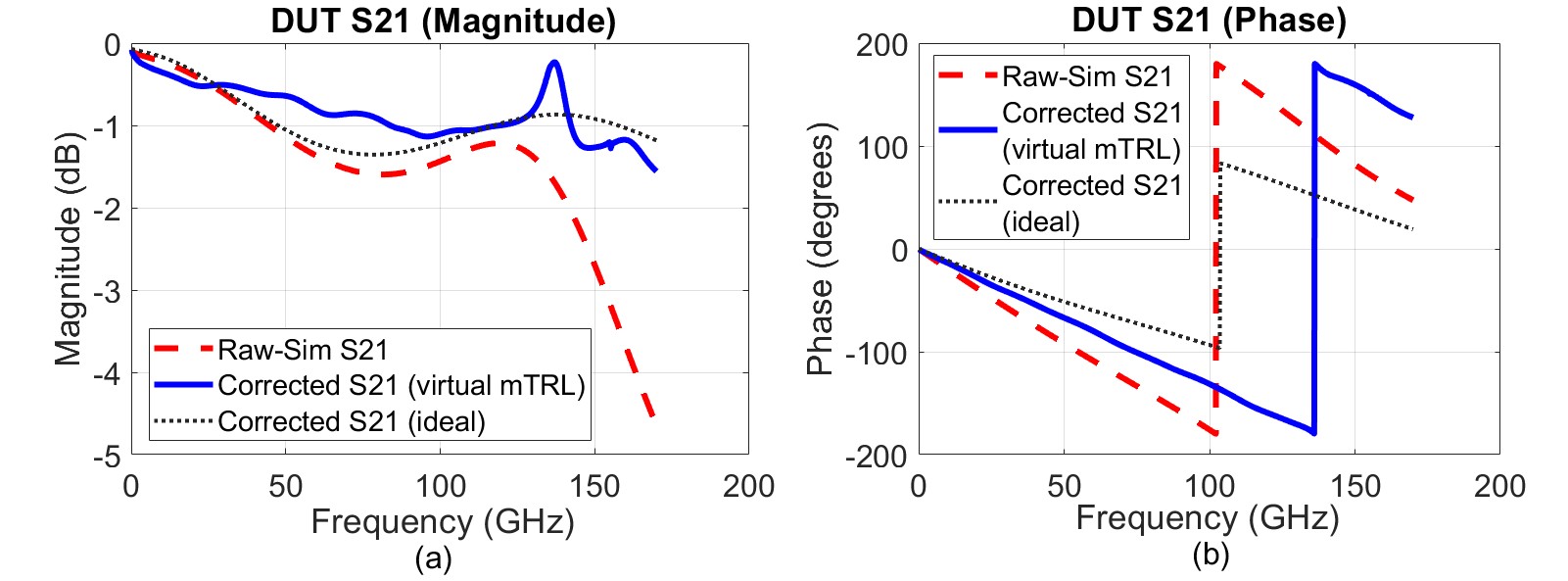}
		\hspace{-5pt}
		\caption{Error correction results on diagonal Thru DUT: (a) $|S_{21}|$ ($<1.5$ dB) ; (b) $\angle S_{21}$. S,O,L standards evaluated w/ virtual mTRL (solid-blue line). Ideal standards used (dotted-black line). Raw uncorrected (dashed-red line)}
		\label{fig:ErrCorr}
		\vspace{-15pt}
	\end{figure}

	Error correction simulations following the SOLR routine \cite{Daniel2005USF} to solve the underlying eight-term error model were completed, employing the simulated s-parameters for the Short, Open, Load, and orthogonal arc Thru standards. A virtual mTRL calibration of the HFSS standard models, based on \cite{Hatab2022arftg}, characterized the standards' s-parameters for the SOLR routine, removing effects of the pads and microstrip lines. The raw-simulated and corrected $|S_{21}|$ magnitude and phase behaviors for the diagonal Thru used as the DUT are presented in Fig. \ref{fig:ErrCorr}. Corrected (solid-blue) traces show improved orthogonal Thru behavior with $|S_{21}|$ magnitude below 1.5 dB. Likewise, $\angle S_{21}$ indicates reduced electrical length, showing a slower linear phase change from DC to 170 GHz, as pads and M9 lines are de-embedded. 
	
	\section{Conclusion}
	This paper highlights two innovative four-port probe stations for simultaneous four-quadrant probing developed independently by MPI Corporation and FormFactor Incorporated. 
	Further, an on-chip four-port SOLR calibration standard is presented for the probe stations that is taped-out on UMC's 28 nm CMOS process. Simulation of the SOLR routine exhibit cal validity up to 125 GHz based on a -15 dB threshold on Load $|S_{11}|$. The four-port probing systems at the NYU \textit{THz Lab} along with the 125 GHz SOLR cal is a critical component for fast and accurate characterization of next-generation multiport circuits and transceivers at mmW and sub-THz without repositioning probes.

	
	

	
	\bibliographystyle{IEEEtran}
	\bibliography{IEEEabrv,IEEEexample}
	
\end{document}